\documentstyle[12pt,aaspp4,natbib]{article}

\def\la{\mathrel{\hbox{\rlap{\hbox{\lower4pt\hbox{$\sim$}}}\hbox{$<$}}}}
\def\ga{\mathrel{\hbox{\rlap{\hbox{\lower4pt\hbox{$\sim$}}}\hbox{$>$}}}}

\slugcomment{Submitted to {\it The Astrophysical Journal}}

\received{2007 xxxxx xx}
\begin{document}

\title{PROMPT Observations of 
the Early-Time Optical Afterglow of GRB 060607A} 

\author{M. Nysewander\altaffilmark{1}\altaffilmark{2}, 
D. E. Reichart\altaffilmark{1}, 
J. A. Crain\altaffilmark{1},
A. Foster\altaffilmark{1},
J. Haislip\altaffilmark{1},
K. Ivarsen\altaffilmark{1},
A. Lacluyze\altaffilmark{1},
A. Trotter\altaffilmark{1}.
}

\altaffiltext{1}{Department of Physics and Astronomy, University of North
Carolina at Chapel Hill, Campus Box 3255, Chapel Hill, NC 27599}
\altaffiltext{2}{Space Telescope Science Institute, 3700 San Martin Drive,
Baltimore, MD, 21218;
mnysewan@stsci.edu}

\begin{abstract}

PROMPT (Panchromatic Robotic Optical Monitoring and Polarimetry Telescopes)
observed the early-time optical afterglow of GRB 060607A and obtained a 
densely sampled
multiwavelength light curve that begins only tens of seconds after the GRB.
Located at Cerro Tololo Inter-American Observatory in Chile, PROMPT is designed
to observe the afterglows of $\gamma$-ray bursts using multiple automated
0.4-m telescopes
that image simultaneously in many filters when the afterglow is bright and
may be highly variable.
The data span the interval from 44 seconds 
after the GRB trigger to 3.3 hours in the $Bgri$ filters.  
We observe an initial 
peak in the light curve at 
approximately three minutes, followed by rebrightenings
peaking around 40 minutes and again at 66 minutes.  Although our data 
overlap with the early \textit{Swift} $\gamma$-ray and x-ray 
light curves, we do not see 
a correlation 
between the optical and high-energy flares.  
We do not find evidence for spectral evolution throughout the 
observations.
We model the variations in the light curves and find that the most likely
cause of the rebrightening episodes is a 
refreshment of the forward shock preceded by a rapidly fading
reverse shock component,
although other explanations are plausible.

\end{abstract}


\section{Introduction}

The general 
behavior of the broadband spectra and light curves of gamma-ray burst 
(GRB) afterglows has been shown to be well described by a jetted, 
relativistically expanding shell 
colliding with an external medium, which has been described as either that 
of a constant density 
\citep{spn98}
or wind-swept environment \citep{cl00}.  
Since the discovery of the first x-ray, optical and radio afterglows over
a decade ago \citep{cfh+97, vgg+97, fkn+97}
astronomers have successfully applied these models 
to derive the conditions of the shock and the 
properties of the surrounding medium.  
In the past decade, the number and quality of 
afterglow follow-up observations have improved dramatically.  The average 
response 
time for a GRB is now minutes instead of days, and 
for this rapid follow up, astronomers regularly use medium 
and large-aperture telescopes 
that can reach deep 
limiting magnitudes.  

However, in nearly all of the most densely sampled, high signal-to-noise 
light curves, significant 
variations are seen super-imposed upon the general behavior.
Three main physical 
scenarios 
have been proposed to explain these features: (1) the ``patchy 
shell" model \citep{kp00}, (2) delayed shocks or energy 
injections from a long-lived central engine \citep{rm98, kp200, sm00}, and
(3) variations in the circumburst density \citep{wl00, dl02}.
Although deviations from the simple power-law decay have been observed
since the beginning of the afterglow era (GRB 970508; \citealt{pmr98}),
the bright afterglows of GRB 021004 and GRB 030329 provide some of the most
densely sampled light curves, and both have been used as test cases for
these three mechanisms.

In the patchy shell model, random 
variations in the energy per unit angle in the 
outflow create the observed bumps in the light curve; the amplitude of these 
bumps is expected to decrease with time.
\citet{npg03} find that this model is slightly preferred over the
other two for GRB 021004 because of the morphology of the fluctuations 
and a consistent value for the electron index between low and high energies.
\citet{dcg+05} also find this to be an acceptable scenario in 
their detailed analysis
of the broadband (optical, near-infrared and millimeter) afterglow
light curve.
However, for GRB 030329
the patchy shell model has been ruled out
because the variations occurred after the jet break, when 
contributions from this effect ought to be negligible \citep{gnp03}.

Energy injections can be described as either delayed 
shocks that are hypothesized to be slower moving shells emitted by the 
central engine that catch up and impact the main shell of the forward
shock as it decelerates \citep{zm02}, or the result of a long-lived
central engine.
In this scenario, the light curve rises briefly as a result of the injection 
after which it resumes its decay with
an index similar to the previous index, but with a new normalization that
reflects the larger energy of the shock.  
Typically, the result of an
energy injection is an overall upward shift of the afterglow
light curve.
Additionally, when the impact occurs, 
it may also produce a bright reverse shock that 
propogates backwards through the ejecta in the comoving frame \citep{np03}.
This scenario was first proposed for
GRB 021004 by \citet{fyk+03} and GRB 030329 by \citet{gnp03}.
For GRB 021004, both \citet{bgj04} and 
\citet{dcg+05} find that the light curves, broadband 
spectral evolution and polarization signature of GRB 021004 can be 
well-modeled with a series of energy injections.  Due to the fact that 
the rebrightening episodes in the light curve of GRB 030329 
appear to be simple re-normalizations, 
both \citet{gnp03} and \citet{hcg06}
find refreshed shocks to be the most natural explanation of the event. 

Shocks resulting from the impact of the fireball upon 
density enhancements in the
surrounding medium can
cause significant rebrightenings in the GRB afterglow.
When the shockwave hits a higher density, the
flux sharply increases before it fades to match a light curve
that is described by the higher density, however, if the density jump is
large enough, it may also produce a reverse shock.
\citet{lrc+02} propose that the variations in the light curve of GRB 021004 are
likely due to moderate changes in density that only slightly modify the
dynamics of the fireball. 
Many groups find that density variations can not account for the
light curve of GRB 030329 due to the increase in flux normalization
after each episode \citep{gnp03, hcg06, uki+04}.

Early multi-color light curves of GRB afterglows are rare.  However, a number
of robotic telescopes have observed afterglows at times early enough to
compare the optical to the high-energy X-ray and $\gamma$-ray emission.
As pointed out by \citet{kp08}, this early-time emission
can be divided into three groups:  (1) The optical emission traces the
high-energy emission (e.g., GRB 041219A; Vestrand et al. 2006); the
optical emission does not trace the high-energy emission (e.g., GRB
990123; Akerlof et al. 1999); and (3) the optical emission has two
components, one that traces the the high-energy emission and a smoother
component, probably the onset of the afterglow, that does not (e.g., GRB
050820A; Vestrand et al. 2006).  GRB 0600607A falls into the second of
these categories (see \S4.1).

PROMPT \citep{rnm+05} has been designed specifically to observe the 
prompt optical and near-infrared 
emission from GRB afterglows simultaneously at multiple 
wavelengths.  The strength of this design is seen in the high-quality, 
densely sampled light curves of GRB 060607A presented in Figure~\ref{0607_lc}.  
Early in the afterglow's lifetime, it evolves on a rapid
time-scale, and 
only with \textit{simultaneous, multiwavelength} observations are 
we able to
properly characterize this phase.
With a data set of this quality we are 
able to ask detailed questions about the nature of the early afterglow 
and explore the possibility of chromatic variations.
In this paper, we present the general properties of the afterglow,
compare the optical, near-infrared and high-energy emission, and
focus on the interpretation of the variability
seen in the light curve.
In \S 2 we review the \textit{Swift} high-energy 
observations and present specific details of the 
PROMPT response.  In \S 3 we fit the standard afterglow and extinction
curve models and discuss the three distinct periods that occur within the first
two hundred minutes.  In \S 4 we compare the optical, near-infrared
and \textit{Swift} x-ray light curves and also discuss the nature
of the observed variations.  We draw conclusions in
\S 5.

\section{Observations}

The \textit{Swift} Burst Alert Telescope (BAT) discovered GRB 060607A
at 05:12:13 UT on June 7$^{th}$, 2006 \citep{zbg+06}.  The burst duration
is T$_{90}$ = 102 seconds (15 -- 350 keV) \citep{sbb+07} and exhibits a
triply-peaked structure \citep{tbb+06}.  Sixty-five seconds after the
initial trigger, the \textit{Swift}'s XRT found a fading x-ray afterglow 
\citep{pgb06},
and at 75 seconds, UVOT began to observe the bright optical afterglow
\citep{obz06}.  \citet{cdm+06} quickly reported a detection in
the near-infrared made by the REM telescope, which began 1.5 minutes after
the trigger \citep{mvm+06}.  \citet{nh06} reported 
$r$ magnitudes that detailed the rise of the early afterglow.  \citet{lvs+06},
began imaging with VLT 7.5 minutes after the BAT trigger and found an
afterglow redshift of z = 3.082.  
\citet{tbb+06} find that using the observed 15 -- 150 keV 
fluence of 2.6 $\pm$ $0.1 \times
10^{-6}$ erg cm$^{-2}$, z = 3.082, 
and standard cosmology ($\Omega_{M}$ = 0.3,
$\Omega_{\Lambda}$ = 0.7, H$_{0}$ = 65), E$_{iso}$ = 1.1 x 10$^{53}$ ergs
in the rest frame 1 -- 1000 keV band.

Four 0.4-m PROMPT telescopes began observing the afterglow of GRB 060607A
in the Bessell $B$ and SDSS $gri$
filters
on June 7$^{th}$, 
2006, 05:12:57 UT, 44 seconds after the initial satellite trigger
and 25 seconds after the GCN notification. 
Observations continued for 5.4 hours under the control of 
Skynet\footnote{Skynet is a dynamic, prioritized 
queue-scheduling system that controls a growing number of 
telescopes, currently spanning North and 
South America; http://skynet.unc.edu}, PROMPT's custom-designed
automation software.   
All images taken 
after 3.3 hours do not yield detections or limiting magnitudes deep or 
significant enough to be included in the analysis.  Table~\ref{0607_tab} 
presents the observations that are plotted in Figure~\ref{0607_lc}.
Figure~\ref{0607_lc} also includes 
the near-infrared $H$-band light curve measured by the
REM telescope \citep{mvm+06}. 

Zero, dark and flat-field calibration images were applied using IRAF's 
CCDPROC package, and, if necessary, images were combined to 
obtain better signal-to-noise.  Point-spread function photometry was applied 
via IRAF's DAOPHOT package to obtain final magnitudes.  
Zeropoints for each image were measured by reobserving the GRB field along with
observing photometric standards 
with PROMPT on the night of June 12, 2006.  Each PROMPT telescope 
uses a 1k x 1k Apogee Alta U47+ CCD, coated with either a midband 
or broadband coating in order to maximize the CCD response in the telescope's 
primary filter set. 
The cameras have fast-readout technology for fast cadence
imaging, with
an inter-exposure deadtime of only 2.5 seconds.

\section{Analysis}

Because of the
five densely sampled distinct light curves, we are able to see two 
significant
rebrightenings, although additional smaller fluctuations 
are suggested by the data.  
Therefore, we model the afterglow as the sum of three peaks, each consisting
of a smoothly broken, rising and falling power-law light curve and a simple
power-law spectrum, where
the spectrum is extinguished by both Milky-Way and source-frame dust and 
absorbed by hydrogen in the source frame and by the
Ly$\alpha$ forest along the line of sight:
\begin{equation}
F_{\nu}(t) = \sum_{n=1} ^3 e^{-\tau_{\nu}^{MW}} e^{-\tau_{\nu(1+z)}^{Ly \alpha}} e^{-\tau_{\nu(1+z)}^{source}} F_{n} \left(\frac{\nu}{\nu_o}\right)^{-\beta_{n}} \left[\left(\frac{t}{t_n}\right)^{-s_{n} \alpha_{n,1}} + \left(\frac{t}{t_n}\right)^{-s_{n} \alpha_{n,2}}\right]^{-1/s_{n}}.
\end{equation}
$\tau_{\nu}^{MW}$ is the Galactic extinction curve model of \citet{ccm89},
$\tau_{\nu(1+z)}^{Ly\alpha}$ is the Ly$\alpha$ forest
absorption model of \citet{r01},
and $\tau_{\nu(1+z)}^{source}$
is the source-frame extinction curve and Lyman-limit absorption model of
\citet{r01}
taking into account the Ly$\alpha$
dampening wing model of \citet{tkk+06}, which is a function of $N_H$.
$\alpha_{1n}$ and $\alpha_{2n}$ are the rising 
($\alpha_{1n} > 0$) and falling ($\alpha_{2n} < 0$)
temporal
indices of the $n^{th}$ peak, $\beta_{n}$ is the spectral index, 
$t_{n}$ is the peak time,
$F_{n}$ is the normalization, $s_{n}$ is the smoothing
parameter, and $\nu_{o}$ is the
effective
frequency of the Sloan $r$ filter.
All magnitudes are converted to fluxes as prescribed by \citet{b79} and
\citet{bb88}.
Since the extinction and absorption models have features that are narrower 
than most photometric bands, we integrate equation (1) against the
appropriate filter transmissivity curve before fitting it to the data.

We fit this model to the data using Bayesian inference 
\citep[e.g.,][]{r01,ltv+01,grb+03,nrp+06}.
The posterior probability distribution is equal to the product
of the prior probability distribution and the likelihood function.  The
likelihood function is given by:
\begin{equation}
{\cal L} = \prod_{i=1}^N \frac{1}{\sqrt{2\pi(\sigma_i^2 + \sigma^2)}} \exp\left\{-\frac{1}{2}\frac{[y(\nu_i,t_i)- y_i]^2}{\sigma_i^2 + \sigma^2}\right\},
\end{equation}
where $N$ is the number of measurements, $y(\nu_{i}, t_{i})$ is the
integration of Equation 1 against the spectral curve of the $i$th
measurement at the time of the $i$th measurement; $y_{i}$ is the $i$th
measurement in units of log spectral flux; $\sigma_{i}$ is the
uncertainty in the $i$th measurement in the same units, and $\sigma$ is a
parameter, sometimes called the slop parameter, that models other,
small sources of uncertainty \citep{r01}.
We account for systematic error in zero point calibrations 
by allowing a magnitude offset for each filter where this offset is
constrained by 
a Gaussian prior of width given by the scatter in the 
relative magnitudes of the calibration stars.
These offsets 
are small and indicate calibration errors less than 0.1 magnitudes for the
PROMPT observations. 
The calibration offset adopted for the REM H-band light curve is $<$0.05
magnitudes.

Many of the parameters of the source-frame extinction curve model and all of
the parameters of the Ly$\alpha$ forest, Galactic extinction
curve, and Ly$\alpha$ damping wing models are constrained a 
priori.  The source-frame extinction curve
model of \citet{r01} is a function of eight parameters: the source-frame
V-band extinction magnitude $A_{V}$, $R_{V} = A_{V}/E(B-V)$, the intercept
$c_{1}$ and slope $c_{2}$ of the linear component of the source-frame UV
extinction curve, the strength $c_{3}$, width $\gamma$ and center $x_{o}$ of
the UV bump component of the extinction curve, and the strength $c_{4}$ of
the FUV excess component of the extinction curve.  The Ly$\alpha$ forest
absorption model of \citet{r01} is a function of a $D_{A}$, the flux
deficit.  
\citet{r01} determines prior probability distributions for $R_{V}$, $c_{1}$, \
$\gamma$, $x_{o}$, and $D_{A}$, which means that the values of these 
parameters can be weighted by fairly narrow distributions, the description
of which depends on other parameters ($c_{2}$ and $z$), a priori.  We adopt
these priors here.  The Galactic extinction curve model of \citet{ccm89}
is a function of $E(B-V)$ = 0.029 mag for this line
of sight \citep{sfd98} and a single parameter,
$R_{V}^{MW}$.  We adopt a prior for this parameter that is log
normally distributed with mean log 3.1 and width 0.1, which 
approximates the distribution of values of this parameter along
random lines of through the Galaxy.
The Ly$\alpha$ dampening wing model 
is a function of the density of neutral hydrogen, $N_{H}$, which
has a prior of $N_H = 7.9^{+3.8}_{-3.5} \times 10^{21}$ cm$^{-2}$
based on the measurement of \citet{pgb06}.

We fit an unconstrained model to the data and then impose relationships between 
the post-peak temporal and spectral indices of 
the first component --
these relationships depend upon the
local environment of the progenitor: a wind-swept (WIND, \citealt{cl00}) or
constant density (ISM, \citealt{spn98}) medium, and the placement of the cooling
break, $\nu_c$, above (BLUE) or below (RED) the observed frequencies.
For the ISM-RED and WIND-RED cases, $\alpha_{1,2} =
(3\beta_1
+ 1)/2 = -(3p - 2)/4$; for the ISM-BLUE case, $\alpha_{1,2} = 3\beta_1/2 = -3(p
- 1)/4$; and
for the WIND-BLUE case, $\alpha_{1,2} = (3\beta_1 - 1)/2 = -(3p - 1)/4$, where
$p$ is the
power-law index of the electron-energy distribution.
In order
to test the spectral variability of the second and third components, only
$\beta_1$ and $\alpha_{1,2}$ are tied together using these relationships.

Using the PROMPT $Bgri$ light curves and the REM $H$-band
light curve \citep{mvm+06},
we find that WIND-BLUE with $p = 2.12^{+0.09}_{-0.07}$
is the most likely scenario, and is consistent with the general fit at the
2.3 $\sigma$ confidence level.  
The ISM-BLUE and ISM/WIND-RED models are
ruled out at the 3.3 and 8.0 $\sigma$ confidence levels, respectively;
only the WIND-BLUE case is consistent with the observed shallow spectral index
$\beta = -0.70$.
Given the observed temporal index of $\alpha_{1,2} = -1.34^{+0.05}_{-0.07}$, for 
ISM-BLUE, $\beta_1 = -0.89^{+0.03}_{-0.05}$, 
and for
ISM/WIND-RED, $\beta_1 = -1.23 \pm$ 0.04.
Often, if a steep spectral slope is observed, degeneracy may exist between 
a steep intrinsic spectral index and the steepening effect of added
extinction, but for GRB 060607A, the observed shallow slope does not
allow for ambiguity.

Although we find a small amount of source-frame extinction, 
$A_{V} = 0.41^{+0.14}_{-0.30}$ mag ($A_{V} > 0$ 
at the 2.6 $\sigma$ credible level),
the parameters for the extinction curve model are not well
constrained.  
$c_{1}$ and $R_{V}$ depend upon $c_{2}$, and 
all are poorly constrained:
$c_{1} = -16.2^{+11.0}_{-7.3}$ and $c_{2} = 5.5^{+2.2}_{-3.3}$, while $R_{V}$
is completely unconstrained.
$c_{3}$ and $c_{4}$ are set to zero
because they are second order effects on 
the global shape of the spectrum.  
Similarly, although the shape of the light curve is well fit by the model
(see Figure~\ref{0607_lc}),
the exact fitting parameter values are again degenerate.  Because of the extreme
slopes involved with the second component, the parameters describing it
are not well-defined.  For the initial onset of the afterglow, 
$F_1 (\mu $Jy) $= 4.32^{+0.04}_{-0.14}$,
$\alpha_{1,1} = 2.46^{+0.23}_{-0.20}$, $\alpha_{1,2} = -1.34^{+0.05}_{-0.07}$, 
$\beta_1 =
-0.56^{+0.03}_{-0.05}$, log $t_1 $(days)$ = -2.70^{+0.01}_{-0.02}$, and $s_1 = 
1.72^{+0.50}_{-0.38}$.
For the second component,
$F_2 (\mu $Jy) $ > 2.83$ (3 $\sigma$), $\alpha_{2,1} > 3.0$ (3 $\sigma$), $\alpha_{2,2} < -3.6$ 
(3 $\sigma$), $\beta_{2} = 
-0.49^{+0.15}_{-0.16}$, log $t_2 $(days)$ = -1.45^{+0.08}_{-0.15}$, and $s_2 <
6.4$ (3 $\sigma$).
For the final component, 
$F_3 (\mu $Jy) $ = 2.53^{+0.20}_{-0.18}$, $\alpha_{3,1} > 0.57$ 
(2 $\sigma$), $\alpha_{3,2} = -1.28^{+0.16}_{-0.20}$,
$\beta_{3} =
-0.16^{+0.17}_{-0.24}$, log $t_3 $(days)$ = -1.36^{+0.04}_{-0.09}$, and $s_3 >
3.29$ (1 $\sigma$).  
All values are cited with 1 $\sigma$ uncertainties, 
however, when only an upper or
lower limit could be placed, the value is cited with the most constraining
limit found (1, 2, or 3 $\sigma$).
Additionally log $N_{H} = 21.69^{+0.24}_{-0.21}$ cm$^{-2}$, 
and the slop parameter 
$\sigma = 0.07 \pm 0.01$ mag.

\section{Discussion}

Three peaks are evident in the prompt
$\gamma$-ray emission observed by the \textit{Swift} BAT: two
overlapping FRED profiles from t$_{0}$ -- 5 to t$_{0}$ + 40 seconds and a third
component at $\approx$100 seconds \citep{tbb+06}.  The 
PROMPT $g$ and $r$ light curves
overlap the third $\gamma$-ray peak, and 
although they are sparsely sampled at this
time, we do not observe any obvious corresponding features.  Similarly,
the peaks in the \textit{Swift}
XRT light curve around 90 and 250 seconds \citep{mvm+06}
do not appear in the PROMPT observations.  Furthermore, neither the initial 
afterglow peak
nor the later variations
at optical frequencies have corresponding features in the x-ray light curve,
although the x-ray light curve at the time of the peaks is not well-sampled.
A comparison of the PROMPT $r$ light curve and the \textit{Swift} XRT
x-ray light curve \citep{ebp+07} is presented in Figure~\ref{0607_xrt}.
Also, within the early optical and near-infrared light curves themselves,
there is no suggestion of chromatic variations.

\subsection{The Early Light Curve}

Both PROMPT and the REM telescope observed the early peak in the
afterglow at three minutes.
\citet{mvm+06} attribute this peak to the 
deceleration of the fireball
and in a detailed analysis, use this peak time
to calculate the Lorentz factor.
Given that the spectral index
should be $+$$1/3$ when $\nu_{opt} < \nu_m$
\citep[e.g.][]{gs02}, the passage of $\nu_m$, the synchrotron peak 
frequency, is ruled
out as a cause for the peak, and hence confirms the result of \citet{mvm+06}.  
Although the pre-peak
spectral index may show signs of evolution, the spectral index is
not positive (see Figure~\ref{0607_lc}), hence $\nu_m < \nu_{opt}$ before the onset
of observations.

The light curve at extreme early times does not appear to be well-fit by the
rising emission, but the discrepancies may be due to lower signal-to-noise of 
the early points.
In the very early light curve, deviations
from the model exist at the 1.8, 0.9 and 1.4$\sigma$
level for the initial $g$, $r$ and $i$ points respectively.
Early variations in the light curve can be interpreted as being due to activity
from the source, however, the early deviations do not
correlate with peaks in the X-ray light curve.  The lower time resolution
of the optical light curves makes it difficult to compare the low and 
high-energy emission.

Spectral evolution before the peak time is similarly difficult to quantify.
The spectral slope measured solely from these three points, centered at 
84 seconds
is poorly constrained: $\beta = -6.5\pm3.3$.
The spectral slope measured from all points before the maximum light 
(during the time period from 65 to 174 seconds) is $\beta = 
-1.18\pm0.52$.
This slope is nearly consistent with the over-all slope of 
$\beta = -0.56^{+0.02}_{-0.05}$ for the first emission component.

\subsection{Comparison with X-Ray Afterglow}

The beginning of the \textit{Swift} era has brought many new questions about
the nature of x-ray afterglows.
X-ray light curves often do not correlate with their optical counterparts,
even when taking into account spectral breaks between the two frequencies.
Possible explanations for the disparity include
a long-lived central engine that continuously
refreshes the forward shock \citep{zfd+06},
two separate components to produce the x-ray and optical afterglows,
or variable microphysical parameters of the shock front \citep{pmb+06}.
Given $p = 2.12^{+0.09}_{-0.07}$ from the optical light curve,
we can examine the correlation between the high and low energy afterglows 
and determine if they
match predictions of the standard model scenarios.

The broadband optical to x-ray spectral flux distribution at twelve minutes
after the burst is presented in 
Figure~\ref{0607_spec}.  The x-ray afterglow flux is taken from \citet{mvm+06} 
and is plotted using the observed x-ray spectral index 
$\beta_x = -0.64 \pm$ 0.07.
$\beta_x$ was measured from the late-time PC data (which range from 600 seconds 
to two days),
and although the early
flares experienced spectral evolution \citep{pgb06}, we assume this value
of $\beta_x$ for the plot.
The SFD is plotted at twelve minutes, which is located immediately 
after the x-ray flares and 
near the break in the x-ray light curve when the temporal index shallows
from $\alpha_{x1} = -1.09 \pm$ 0.04 to $\alpha_{x2} = -0.45 \pm$ 0.03 
\citep{pgb06}.
The broadband spectral flux distribution 
indicates that, at this time, the low and high energy light curves may be
correlated.  
Although uncertain given the temporal scaling and insecure spectral slope, 
Figure~\ref{0607_spec} shows the 
extrapolation of the flux from the x-ray to optical band, which suggests that 
before the break in the x-ray light curve, one underlying mechanism may 
be producing both high and low-energy emission.

However, the best fit model to the optical and near-infrared afterglow 
is one with $\nu_c$ blueward of
optical frequencies ($\nu_c > \nu_{opt}$ at the 8.0 $\sigma$ credible level).  
Contrary to a constant density medium, for the synchrotron spectrum of a
shock expanding 
in a wind-swept environment, $\nu_c$
moves upwards through frequency space.
At twelve minutes, although $\nu_c > \nu_{opt}$, $\nu_c$ has not likely passed
above x-ray frequencies, hence the lack of a break in Figure~\ref{0607_spec}
would be inconsistent with our best fit.  

Indeed, the difference in temporal
slope at this time ($\alpha_{opt} = -1.34^{+0.05}_{-0.07}$ vs. $\alpha_{x1} =
-1.09 \pm 0.04$ or $\alpha_{x2} = -0.45 \pm 0.03$) 
indicates that there is a break between 
the high and low frequencies.
Therefore, it is possible that $\beta_x$ underwent a change in slope 
between twelve
minutes and when it was measured, and our assumed spectral slope plotted in
Figure~\ref{0607_spec} is incorrect.
In the WIND-BLUE fit: $p = 2.12^{+0.09}_{-0.07}$,
$\alpha_{opt} = -1.34^{+0.05}_{-0.07}$ and  
$\beta_{opt} = -0.56^{+0.03}_{-0.05}$.  If $\nu_{opt} < \nu_c < \nu_x$, 
then the x-ray data will be described by the WIND-RED case and 
the predicted values for the temporal and spectral x-ray slopes are:
$\alpha_{x, W-R} = -1.09^{+0.05}_{-0.07}$ and 
$\beta_{x,W-R} = -1.06^{+0.03}_{-0.05}$.
Note that the predicted value for the x-ray temporal slope matches
the observed early x-ray slope of $\alpha_{x, W-R} = \alpha_x = -1.09$, 
hence the pre-break ($t < 12$ minutes) period
of the x-ray light may be produced by the 
the same synchrotron component
as that of the optical light curve.
The break to the shallower, $\alpha_{x2} = -0.45 \pm 0.03$, slope 
would then correspond to the onset of a second phase of emission not
associated with the optical afterglow.

\subsection{Modeling the Variations}

Two clear achromatic
rebrightening episodes are observed in the optical and near-infrared
light curves at 40 and 66 minutes.
Here we consider possible scenarios for these peaks: the passage of a 
spectral break, the patchy-shell model, density variations, refreshed
shocks, and associated reverse shocks.
We do not expect that the variations in the light curve will be due to more
than one scenario, as it would be unlikely for multiple conditions
to be satisfied \citep{npg03}.

At the time near the peaks of the latter two components, the 
x-ray light curve is noisy
and not well-sampled, however it appears to flatten before
the beginning of the first optical rebrightening and
remains shallow until the end of the PROMPT light curve before dropping
suddenly at $\approx$4 hours (see Figure~\ref{0607_xrt}).  
Because $\nu_c$ is likely 
between the optical and x-ray bands,
the lack of features in the high-energy emission may suggest that the
episodes are due to density variations in the circumburst medium:
variations that depend upon changes in energy
would produce variations both above and below $\nu_c$.
However, without a clear understanding of the nature of the x-ray light
curve and the relationship between it and the optical, this claim is
questionable.
If, as suggested in Section \S4.1, the nearly flat phase of the 
x-ray emission is not produced by the same mechanism as that of the optical,
then evidence for variations above $\nu_c$ may be masked by the flux from 
a second
component.

\subsubsection{Spectral Evolution}

We first consider whether the peaks are due to the passage of a break --
although the indices are not generally well constrained,
they may suggest spectral evolution:
$\beta_1 = -0.56^{+0.03}_{-0.05}$, $\beta_2 = -0.49^{+0.15}_{-0.16}$ and 
$\beta_3 = -0.16^{+0.17}_{-0.24}$.
Deviations from our simple models of breaks in the electron distribution 
might influence the synchrotron spectrum and light curve in unexpected
ways. 
A model with no spectral evolution, where we impose that
$\beta_1 = \beta_3$ differs from the WIND-BLUE model at only the 1.9 $\sigma$ 
confidence level.
The observed change in spectral index ($\beta_3 = \beta_1 + 0.40 \pm 0.18$), 
from a moderate to 
shallow slope, is roughly consistent with the passage of 
$\nu_c$ from low to high
frequencies.  
However, the WIND and ISM-RED models are ruled out for the early data at the
8.0 $\sigma$ confidence level, and hence we do not find this to be
a likely scenario.

\subsubsection{Patchy Shells}

The patchy-shell model, where the observed flux depends upon the
angular structure of the jet, has variations with a
timescale
longer than the observed time, $\Delta t > t$, and
more specifically, has decay times, $t_{decay} \approx t$ \citep{no04}.
At the end of the fluctuations, the light curve will fade to match the
extrapolation of the power-law light curve before the episode occurred.
The timescale for the first variation is short, $\Delta t \approx 27$
min, but the peak of the variation is at 33 min,
therefore the first relation is marginally satisfied.
However, the first peak decays rapidly, $t_{decay} \approx 15$ min, 
which is too extreme to
be expected from 
the timescale of the observations.
The second variation may satisfy both of these 
relationships;
however, the post-peak decay index does not indicate that 
the light curve is fading to its pre-episode flux.
Therefore, because the first variation is not consistent with the
predicted constraints and the decay of the second variation matches
the pre-episode index, the patchy shell model is
also unlikely.

\subsubsection{Density Variations}

Variations due to density enhancements in the circumburst environment  
can inflate the flux of the light curve that then settles into the 
afterglow solution describing the new environment. 
Generally, for a single density fluctuation, the light curve will asymptotically
relax to the 
pre-episode brightness.
However, if the shock impacts a shell after which the density increases
permanently,
then it is possible for the light curve to appear to renormalize.
In addition to this, as in the patchy-shell scenario, it is difficult for 
even a very sharp fluctuation in the density to produce
short time-scale fluctations 
in the light curve \citep{np03} with $\Delta t < t$.
Therefore, we do not find it likely 
that the first of the two variations is produced 
via this scenario.

However, if the jump in density is high enough ($\sim$21, \citealt{dl02}), then 
a reverse shock may form in addition to the enhancement of the forward shock. 
Due to the rapidly fading nature of the second feature, we find this to
be a plausible explanation for the second episode.
In this scenario, though, the apparent renormalization after the second
variation is a coincidence not predicted by the model, 
and relies upon conditions specific to the
circumburst density profile.

\subsubsection{Energy Injection}

If the bumps are due to two episodes of energy
injection, either by the continuing activity of the progenitor or 
by late internal shocks resulting from slower shells
impacting the
main forward shock,
then the fading temporal ($\alpha_{1,2} = \alpha_{2,2} =
\alpha_{3,2}$) and spectral ($\beta_1 = \beta_2 = \beta_3$) indices will 
remain nearly constant.
The energy injection will raise the flux of the light curve and
although the post-shock indices will reflect the new shock
conditions,
they should vary little from the previous state.
Due to the steepness of $\alpha_4$ this model is ruled out at the 
7.5 $\sigma$ confidence level.

However, as in the case of density variations, it is possible that
internal shocks can cause a bright
reverse shock to form
and be observed
as a bright, rapidly fading flare in the GRB afterglow. 
In this case, the first and third variations will have nearly the same indices 
($\alpha_{1,2} = \alpha_{3,2}$;
$\beta_1 = \beta_3$) and the second variation will fade rapidly 
($\alpha_{2,2} \geq 2$).
Quantitatively, this model is consistent with the general model at the
1.9 $\sigma$ uncertainty level.
Qualitatively, this is precisely what we see: the second episode at 
66 minutes is a simple renormalization
after which the afterglows fades with the same initial slope, whereas the first
clearly fades more steeply. 
In fact, focusing on the pre- and post-episode slopes, a model 
with $\alpha_{1,2} = \alpha_{3,2}$ is consistent with a model
where the parameters are unconstrained at the 0.4 $\sigma$ confidence level.
Therefore, we find that a refreshed shock that was preceeded by a bright
reverse shock is the most likely cause of the two rebrightening episodes.
The observations are consistent with density variations, however, the
renormalization required in that circumstance would be merely 
coincidental, rather
than a prediction of the model.

\section{Conclusions}

We have presented the early-time PROMPT afterglow light curve for GRB 060607A,
fit a model to the data and found that the progenitor likely exploded
into a wind-swept medium and that the cooling frequency of the synchrotron
spectrum is above the optical frequencies before the beginning of the
PROMPT observations.
Observing correlations between peaks in
different energy ranges is a key tool for probing the underlying physics
of the central engine and its interaction with its environment.
Although the x-ray spectrum exhibited intense flares at early times,
these flares are not present in the optical light curves.
Likewise, optical flares at later times are not observed in
the x-ray light curve.  
In the PROMPT optical light curve we see an initial peak 
at 2.9 minutes, and later, 
two clear rebrightening episodes at 40 and 66 minutes that are not
observed at high energies.
We do not observe spectral evolution throughout the PROMPT dataset and 
constrain $\nu_m$ to be below and $\nu_c$ to be above optical 
frequencies by the onset of observations.

Many authors have cited the need for observations at both low and high energy
in order to distinguish between 
variations due to energy or density, and in this case the lack of features
in the high-energy light curve would indicate that the fluctations are due
to density.
However, without a clear understanding of the
mechanisms that produce the light curves, we cannot definitively relate
them.
We find that the most likely scenario for the observed fluctuations
is that of energy injection to the forward shock, either through a delayed 
shock or continuing activity of the central engine, which 
is strong enough to produce
a bright reverse shock that sweeps through the relativistic
material behind the main forward shock.
In the case of GRB 060607, although energy injection is the favored 
scenario,
the constraints imposed by the light curves are not conclusive proof.

As efforts to observe the afterglows of $\gamma$-ray bursts have 
strengthened, data
sets grow more robust.
Clearly, if variations are ubiquitous, the analysis of a poorly sampled
single light curve is difficult to trust.
The results obtained via 
modeling such a light curve 
that has undergone episodes of rebrightening will likely deviate from its true
behavior, and the steepness of
temporal decays
will be systematically underestimated.
PROMPT is a powerful tool that can produce densely sampled early-time
light curves when afterglows are bright and more 
likely to undergo rapid variations.
By observing simultaneously in multiple filters, we can test the early-time
afterglow for chromatic variations, measure spectral indices, and 
obtain multiple, independent
light curves.  Because of these factors, PROMPT will play a strong 
role in future afterglow follow-up 
studies.

\acknowledgements
DER very gratefully acknowledges support from NSF's MRI, CAREER, AAG, 
and PREST programs, NASA's APRA, Swift GI, and IDEAS programs, NC 
Space Grant's NIP program, and especially Leonard Goodman and Henry Cox. 
We gratefully acknowledge Don Smith for useful comments and collaboration
on the PROMPT project, and also T. Brennan,
M. Schubel and J. Styblova for their work as new members of the 
UNC GRB Follow-Up Group.
This work made use of data supplied by the UK Swift Science Data Centre at 
the University of Leicester.

\clearpage

\begin{deluxetable}{lcccc}
\tablecolumns{5}
\tablewidth{0pt}
\tablecaption{PROMPT Observations of the Afterglow of GRB 060607A}
\tablehead{\colhead{Mean Time (UT)} & \colhead{Exp. Time} & \colhead{Mean $\Delta$t (hr)} & \colhead{Filter} 
& \colhead{Magnitude}}
\startdata
Jun 7	5:13:38	&	15	s	&	0.0235	&	$B$	&	$>$16.61	\\
Jun 7	5:14:09	&	20	s	&	0.0322	&	$B$	&	16.44	$\pm$	0.10	\\
Jun 7	5:14:40	&	20	s	&	0.0408	&	$B$	&	15.69	$\pm$	0.07	\\
Jun 7	5:15:11	&	20	s	&	0.0494	&	$B$	&	15.47	$\pm$	0.06	\\
Jun 7	5:15:42	&	20	s	&	0.0581	&	$B$	&	15.63	$\pm$	0.07	\\
Jun 7	5:16:25	&	40	s	&	0.0700	&	$B$	&	15.59	$\pm$	0.04	\\
Jun 7	5:17:15	&	40	s	&	0.0839	&	$B$	&	15.84	$\pm$	0.05	\\
Jun 7	5:18:06	&	40	s	&	0.0981	&	$B$	&	16.03	$\pm$	0.06	\\
Jun 7	5:18:56	&	40	s	&	0.1119	&	$B$	&	16.21	$\pm$	0.07	\\
Jun 7	5:20:07	&	80	s	&	0.1317	&	$B$	&	16.51	$\pm$	0.09	\\
Jun 7	5:21:45	&	80	s	&	0.1589	&	$B$	&	16.55	$\pm$	0.05	\\
Jun 7	5:23:23	&	80	s	&	0.1861	&	$B$	&	16.86	$\pm$	0.06	\\
Jun 7	5:25:01	&	80	s	&	0.2133	&	$B$	&	17.16	$\pm$	0.11	\\
Jun 7	5:26:36	&	80	s	&	0.2397	&	$B$	&	17.25	$\pm$	0.12	\\
Jun 7	5:28:53	&	160	s	&	0.2778	&	$B$	&	17.60	$\pm$	0.09	\\
Jun 7	5:32:44	&	240	s	&	0.3420	&	$B$	&	17.77	$\pm$	0.09	\\
Jun 7	5:37:26	&	240	s	&	0.4203	&	$B$	&	17.89	$\pm$	0.08	\\
Jun 7	5:42:04	&	240	s	&	0.4974	&	$B$	&	17.85	$\pm$	0.07	\\
Jun 7	5:46:36	&	240	s	&	0.5730	&	$B$	&	17.75	$\pm$	0.08	\\
Jun 7	5:51:50	&	320	s	&	0.6601	&	$B$	&	17.98	$\pm$	0.07	\\
Jun 7	6:00:52	&	640	s	&	0.8110	&	$B$	&	18.56	$\pm$	0.09	\\
Jun 7	6:14:19	&	800	s	&	1.0350	&	$B$	&	18.61	$\pm$	0.11	\\
Jun 7	6:28:11	&	640	s	&	1.2662	&	$B$	&	18.91	$\pm$	0.11	\\
Jun 7	7:09:32	&	960	s	&	1.9553	&	$B$	&	19.53	$\pm$	0.19	\\
Jun 7	7:48:25	&	1200	s	&	2.6033	&	$B$	&	19.98	$\pm$	0.18	\\
Jun 7	8:18:06	&	1360	s	&	3.0981	&	$B$	&	19.75	$\pm$	0.14	\\
Jun 7	5:13:20	&	5	s	&	0.0185	&	$g$	&	16.63	$\pm$	0.15	\\
Jun 7	5:13:46	&	10	s	&	0.0258	&	$g$	&	16.11	$\pm$	0.10	\\
Jun 7	5:14:46	&	10	s	&	0.0425	&	$g$	&	15.04	$\pm$	0.03	\\
\enddata
\tablenotetext{a}{Upper limits are 3$\sigma$.}
\label{0607_tab}
\end{deluxetable}

\begin{deluxetable}{lcccc}
\tablecolumns{5}
\tablewidth{0pt}
\tablecaption{PROMPT Observations of the Afterglow of GRB 060607A (con't)}
\tablehead{\colhead{Mean Time (UT)} & \colhead{Exp. Time} & \colhead{Mean $\Delta$t (hr)} & \colhead{Filter}
& \colhead{Magnitude}}
\startdata
Jun 7   5:15:48 &       20      s       &       0.0597  &       $g$      &
15.09   $\pm$   0.03    \\
Jun 7   5:17:17 &       20      s       &       0.0844  &       $g$      &
15.37   $\pm$   0.02    \\
Jun 7	5:19:03	&	40	s	&	0.1139	&	$g$	&	15.82	$\pm$	0.02	\\
Jun 7	5:21:55	&	80	s	&	0.1617	&	$g$	&	16.10	$\pm$	0.01	\\
Jun 7	5:25:06	&	80	s	&	0.2147	&	$g$	&	16.67	$\pm$	0.03	\\
Jun 7	5:29:45	&	80	s	&	0.2922	&	$g$	&	17.20	$\pm$	0.03	\\
Jun 7	5:31:17	&	80	s	&	0.3178	&	$g$	&	17.30	$\pm$	0.05	\\
Jun 7	5:35:57	&	80	s	&	0.3956	&	$g$	&	17.38	$\pm$	0.04	\\
Jun 7	5:37:29	&	80	s	&	0.4211	&	$g$	&	17.31	$\pm$	0.04	\\
Jun 7	5:42:11	&	80	s	&	0.4994	&	$g$	&	17.27	$\pm$	0.04	\\
Jun 7	5:43:42	&	80	s	&	0.5247	&	$g$	&	17.36	$\pm$	0.04	\\
Jun 7	5:45:17	&	80	s	&	0.5511	&	$g$	&	17.33	$\pm$	0.04	\\
Jun 7	5:46:27	&	40	s	&	0.5706	&	$g$	&	17.37	$\pm$	0.07	\\
Jun 7	5:54:12	&	80	s	&	0.6997	&	$g$	&	18.04	$\pm$	0.06	\\
Jun 7	5:55:42	&	80	s	&	0.7247	&	$g$	&	18.04	$\pm$	0.07	\\
Jun 7	5:57:13	&	80	s	&	0.7500	&	$g$	&	18.05	$\pm$	0.05	\\
Jun 7	5:58:24	&	40	s	&	0.7697	&	$g$	&	18.15	$\pm$	0.07	\\
Jun 7	6:08:25	&	280	s	&	0.9367	&	$g$	&	18.52	$\pm$	0.17	\\
Jun 7	6:26:04	&	240	s	&	1.2307	&	$g$	&	18.41	$\pm$	0.08	\\
Jun 7	6:30:39	&	240	s	&	1.3071	&	$g$	&	18.41	$\pm$	0.05	\\
Jun 7	7:05:40	&	240	s	&	1.8907	&	$g$	&	19.20	$\pm$	0.17	\\
Jun 7	7:42:28	&	640	s	&	2.5042	&	$g$	&	19.48	$\pm$	0.06	\\
Jun 7	8:22:50	&	800	s	&	3.1768	&	$g$	&	19.74	$\pm$	0.08	\\
Jun 7	5:13:18	&	10	s	&	0.0181	&	$r$	&	15.98	$\pm$	0.31	\\
Jun 7	5:13:44	&	10	s	&	0.0253	&	$r$	&	15.75	$\pm$	0.16	\\
Jun 7	5:14:43	&	20	s	&	0.0417	&	$r$	&	14.34	$\pm$	0.03	\\
Jun 7	5:15:45	&	20	s	&	0.0589	&	$r$	&	14.32	$\pm$	0.03	\\
Jun 7	5:17:19	&	40	s	&	0.0850	&	$r$	&	14.68	$\pm$	0.02	\\
Jun 7	5:19:01	&	40	s	&	0.1133	&	$r$	&	15.11	$\pm$	0.02	\\
\enddata
\end{deluxetable}

\begin{deluxetable}{lcccc}
\tablecolumns{5}
\tablewidth{0pt}
\tablecaption{PROMPT Observations of the Afterglow of GRB 060607A (con't)}
\tablehead{\colhead{Mean Time (UT)} & \colhead{Exp. Time} & \colhead{Mean $\Delta$t (hr)} & \colhead{Filter}
& \colhead{Magnitude}}
\startdata
Jun 7   5:21:47 &       80      s       &       0.1594  &       $r$      &
15.47   $\pm$   0.02    \\
Jun 7   5:25:03 &       80      s       &       0.2139  &       $r$      &
16.04   $\pm$   0.03    \\
Jun 7   5:29:43 &       80      s       &       0.2917  &       $r$     &
16.60   $\pm$   0.05    \\
Jun 7   5:31:12 &       80      s       &       0.3164  &       $r$      &
16.57   $\pm$   0.05    \\
Jun 7	5:35:56	&	80	s	&	0.3953	&	$r$	&	16.57	$\pm$	0.04	\\
Jun 7	5:37:26	&	80	s	&	0.4203	&	$r$	&	16.75	$\pm$	0.08	\\
Jun 7	5:42:09	&	80	s	&	0.4989	&	$r$	&	16.69	$\pm$	0.07	\\
Jun 7	5:43:38	&	80	s	&	0.5236	&	$r$	&	16.60	$\pm$	0.06	\\
Jun 7	5:45:11	&	80	s	&	0.5494	&	$r$	&	16.64	$\pm$	0.08	\\
Jun 7	5:46:20	&	40	s	&	0.5686	&	$r$	&	16.78	$\pm$	0.12	\\
Jun 7	5:54:10	&	80	s	&	0.6992	&	$r$	&	17.25	$\pm$	0.08	\\
Jun 7	5:55:39	&	80	s	&	0.7239	&	$r$	&	17.37	$\pm$	0.08	\\
Jun 7	5:57:09	&	80	s	&	0.7489	&	$r$	&	17.37	$\pm$	0.10	\\
Jun 7	5:58:38	&	80	s	&	0.7736	&	$r$	&	17.54	$\pm$	0.12	\\
Jun 7	6:08:24	&	280	s	&	0.9365	&	$r$	&	17.65	$\pm$	0.11	\\
Jun 7	6:28:15	&	480	s	&	1.2672	&	$r$	&	17.69	$\pm$	0.06	\\
Jun 7	7:05:04	&	320	s	&	1.8810	&	$r$	&	18.44	$\pm$	0.08	\\
Jun 7	7:41:40	&	720	s	&	2.4909	&	$r$	&	18.77	$\pm$	0.09	\\
Jun 7	8:23:22	&	880	s	&	3.1859	&	$r$	&	19.04	$\pm$	0.15	\\
Jun 7	5:14:12	&	20	s	&	0.0331	&	$i$	&	14.75	$\pm$	0.05	\\
Jun 7	5:15:14	&	20	s	&	0.0503	&	$i$	&	14.05	$\pm$	0.04	\\
Jun 7	5:16:27	&	40	s	&	0.0706	&	$i$	&	14.20	$\pm$	0.04	\\
Jun 7	5:18:10	&	40	s	&	0.0992	&	$i$	&	14.76	$\pm$	0.05	\\
Jun 7	5:20:13	&	80	s	&	0.1333	&	$i$	&	15.01	$\pm$	0.04	\\
Jun 7	5:23:25	&	80	s	&	0.1867	&	$i$	&	15.61	$\pm$	0.05	\\
Jun 7	5:26:38	&	80	s	&	0.2403	&	$i$	&	15.99	$\pm$	0.07	\\
Jun 7	5:28:07	&	80	s	&	0.2650	&	$i$	&	16.23	$\pm$	0.08	\\
Jun 7	5:32:49	&	80	s	&	0.3433	&	$i$	&	16.34	$\pm$	0.10	\\
Jun 7	5:34:18	&	80	s	&	0.3681	&	$i$	&	16.43	$\pm$	0.08	\\
\enddata
\end{deluxetable}

\begin{deluxetable}{lcccc}
\tablecolumns{5}
\tablewidth{0pt}
\tablecaption{PROMPT Observations of the Afterglow of GRB 060607A (con't)}
\tablehead{\colhead{Mean Time (UT)} & \colhead{Exp. Time} & \colhead{Mean $\Delta$t (hr)} & \colhead{Filter}
& \colhead{Magnitude}}
\startdata
Jun 7	5:39:02	&	80	s	&	0.4469	&	$i$	&	16.48	$\pm$	0.08	\\
Jun 7	5:40:31	&	80	s	&	0.4717	&	$i$	&	16.53	$\pm$	0.07	\\
Jun 7	5:48:08	&	80	s	&	0.5986	&	$i$	&	16.68	$\pm$	0.09	\\
Jun 7	5:49:37	&	80	s	&	0.6233	&	$i$	&	16.60	$\pm$	0.11	\\
Jun 7	5:51:08	&	80	s	&	0.6486	&	$i$	&	16.79	$\pm$	0.10	\\
Jun 7	6:00:27	&	400	s	&	0.8038	&	$i$	&	17.33	$\pm$	0.07	\\
Jun 7	6:15:52	&	480	s	&	1.0609	&	$i$	&	17.46	$\pm$	0.07	\\
Jun 7	7:13:22	&	560	s	&	2.0190	&	$i$	&	18.32	$\pm$	0.09	\\
Jun 7	8:03:47	&	720	s	&	2.8594	&	$i$	&	18.94	$\pm$	0.19	\\
\enddata
\end{deluxetable}

\clearpage

\figcaption[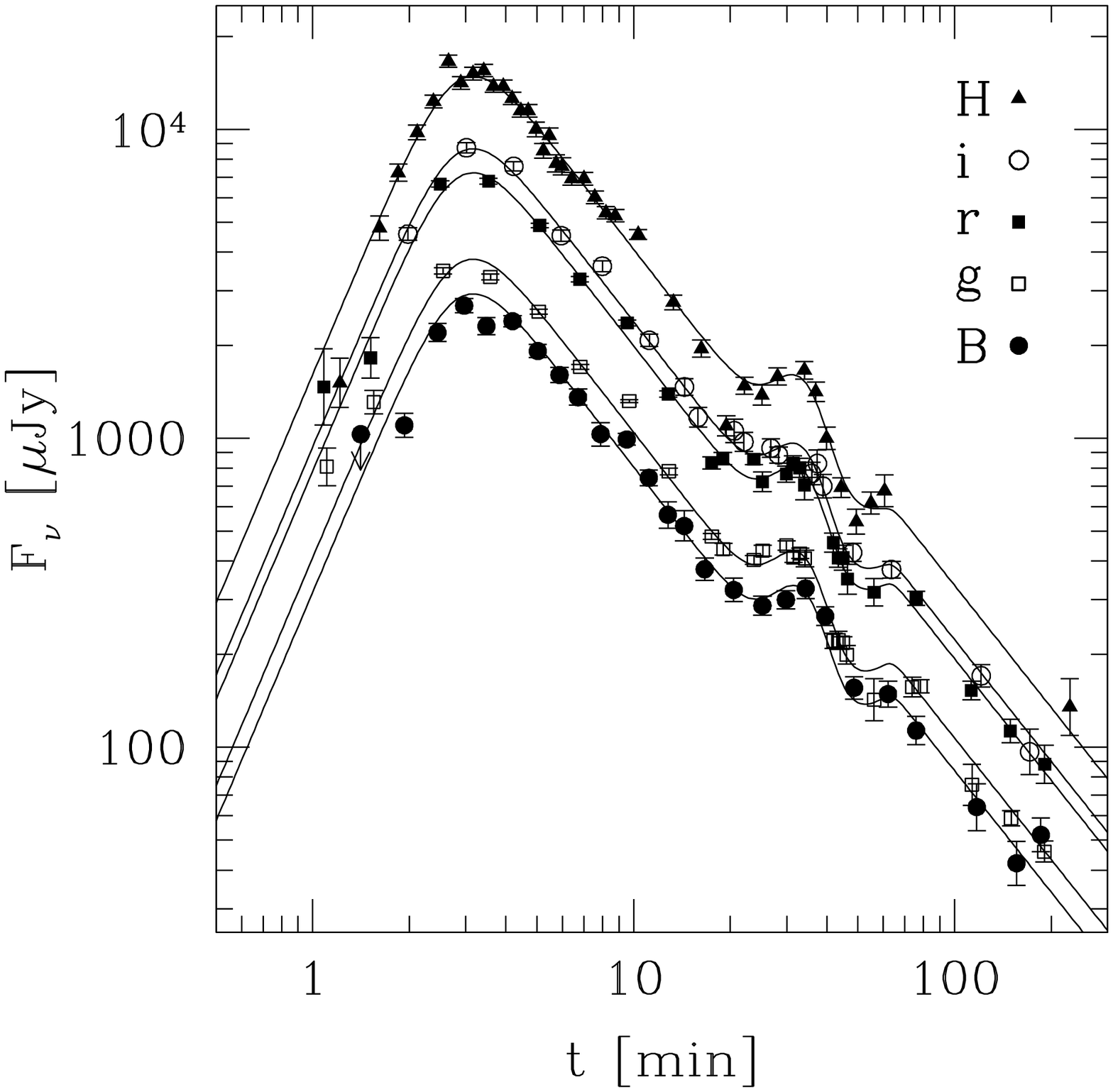]{The $Bgri$ afterglow of GRB 060607A from 44 seconds 
to 3.2 hours taken with the PROMPT telescopes 
along with the $H$-band REM
light curve \citep{mvm+06}.  The solid lines present the 
best fit to the initial peak
at three minutes and two 
variations that occur at 42 and 64 minutes.
\label{0607_lc}}

\figcaption[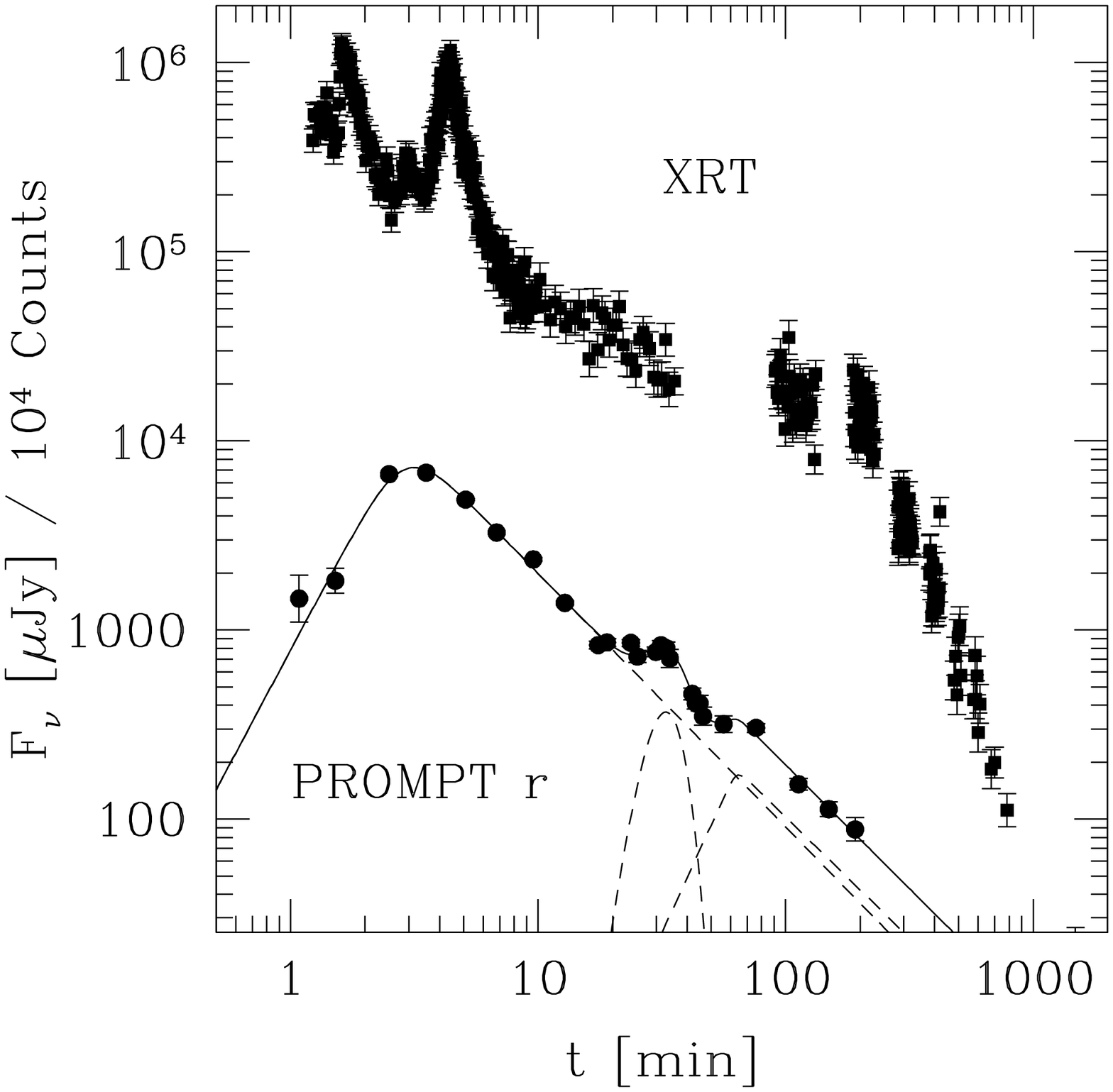]{
The PROMPT $r$-band light curve plotted with the XRT x-ray afterglow
scaled to units of 10$^{4}$ counts.  Note that the time of the
variations in the optical correspond to a lapse in x-ray observations.
However, the x-ray light curve appears to flatten around this time and hence
does not undergo the same fluctuating
behavior as the optical light curve.
\label{0607_xrt}}

\figcaption[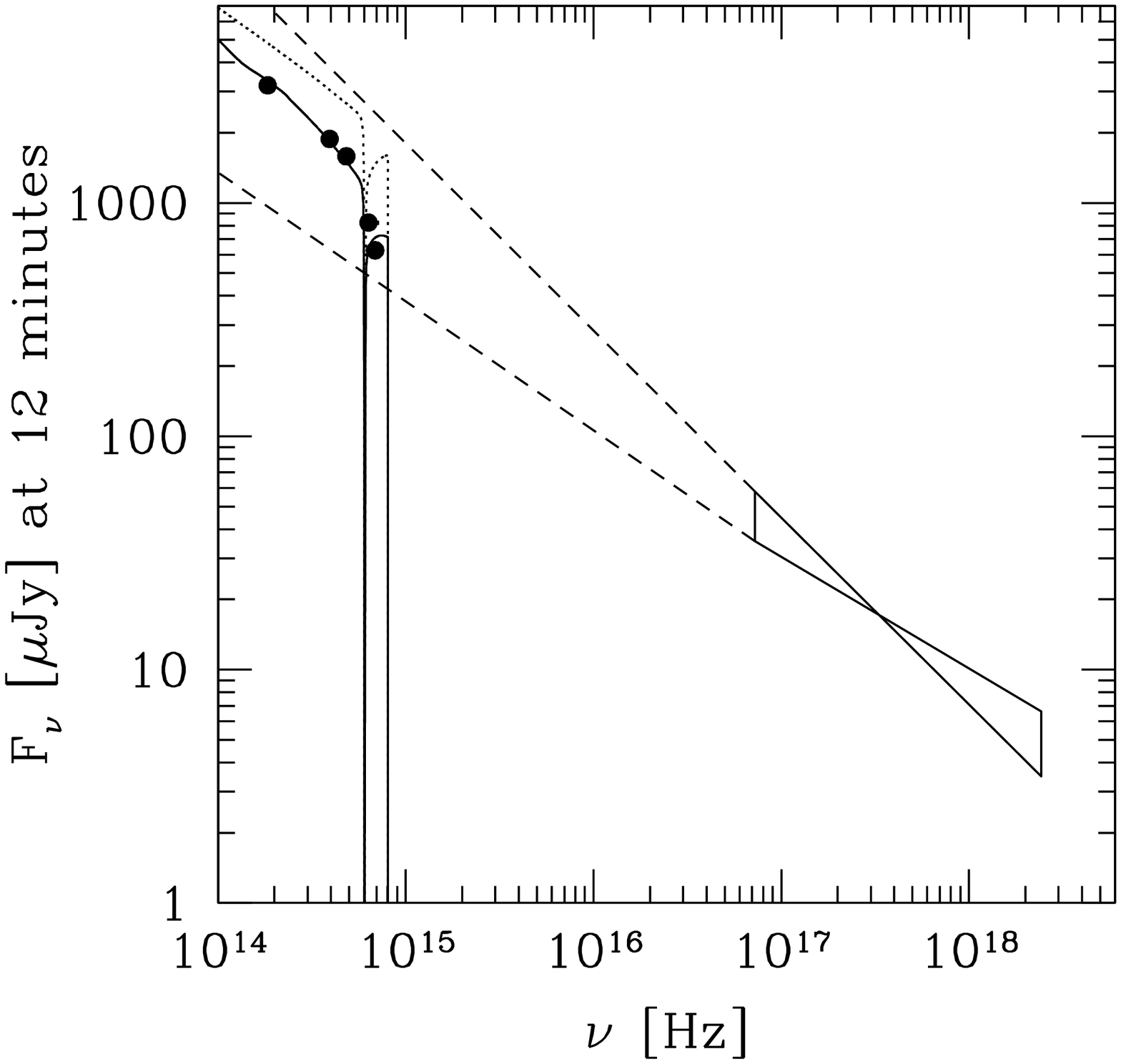]{The broadband spectral flux distribution
of the afterglow of 
GRB 060607A plotted at twelve minutes.  The solid lines are the fit to the
data; the dotted lines are the unextinguished model.  The high energy x-ray
flux is adopted from \citet{mvm+06}, and the plotted spectral index is 
taken from \citet{pgb06}.
\label{0607_spec}}

\clearpage

\setcounter{figure}{0}

\begin{figure}[tb]
\plotone{f1.eps}
\end{figure}

\begin{figure}[tb]
\plotone{f2.eps}
\end{figure}

\begin{figure}[tb]
\plotone{f3.eps}
\end{figure}

\clearpage

\bibliographystyle{apj}                 
\bibliography{ms} 	                

\end{document}